\documentclass{vgtc}                          

\ifpdf
  \pdfoutput=1\relax                   
  \usepackage{graphicx}                
  \DeclareGraphicsExtensions{.pdf,.png,.jpg,.jpeg} 
\else
  \usepackage{graphicx}                
  \DeclareGraphicsExtensions{.eps}     
16

\fi%

\graphicspath{{figures/}{pictures/}{images/}{./}} 

\usepackage{microtype}                 
\PassOptionsToPackage{warn}{textcomp}  
\usepackage{textcomp}                  
\usepackage{mathptmx}                  
\usepackage{times}                     
\usepackage{cite}                      
\usepackage{tabu}                      
\usepackage{booktabs}                  
\usepackage{tikz}

\usepackage{fancyhdr}

\onlineid{0}

\vgtccategory{Research}

\vgtcinsertpkg

\title{Developing Visualisations to Enhance an Insider Threat Product: A Case Study}

\author{Martin Graham\thanks{e-mail: m.graham@napier.ac.uk}\\ %
\parbox{1.4in}{\scriptsize \centering Edinburgh Napier University \\ \& FortiNet Ltd}
\and Robert Kukla\thanks{e-mail: r.kukla@napier.ac.uk}\\ %
\scriptsize Edinburgh Napier University %
\and Oleksii Mandrychenko\thanks{e-mail: omandrychenko@fortinet.com}\\ %
\scriptsize FortiNet Ltd %
\and Darren Hart\thanks{e-mail: dhart@fortinet.com}\\ %
\scriptsize FortiNet Ltd %
\and Jessie Kennedy\thanks{e-mail: j.kennedy@napier.ac.uk}\\ %
\scriptsize Edinburgh Napier University} %

\teaser{
  \centering
    \begin{tikzpicture}
        \node[anchor=south west,inner sep=0] at (0,0) {\fbox{\includegraphics[width=\linewidth]{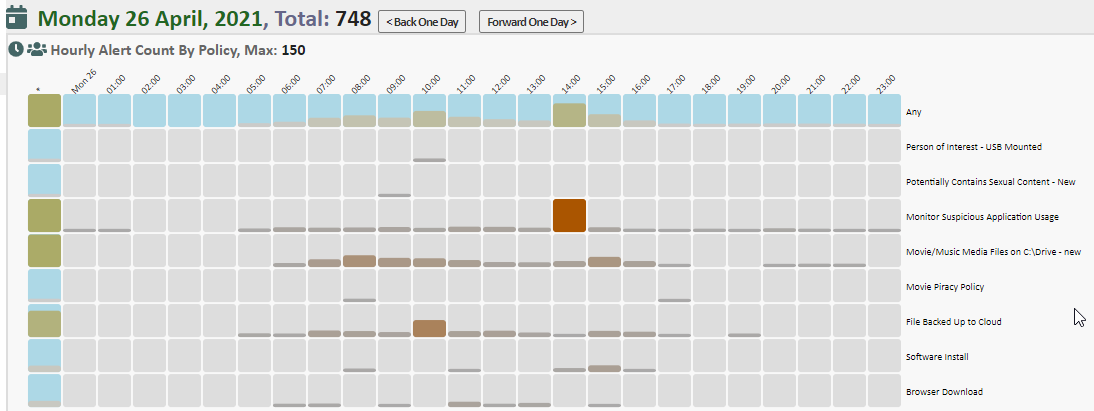}}};
    \end{tikzpicture}
    \caption{Detail of a grid sub-view component of the insider threat visualisation. 24 hours of alerts subdivided by the policy violations that caused the alerts. Peak activity was at 2pm and primarily caused by the Monitor Suspicious Application Usage policy.} 
    \label{fig:teaser}
}

\abstract{
This paper describes the process of developing data visualisations to enhance a commercial software platform for combating insider threat, whose existing UI, while perfectly functional, was limited in its ability to allow analysts to easily spot the patterns and outliers that visualisation naturally reveals. We describe the design and development process, proceeding from initial tasks/requirements gathering, understanding the platform's data formats, the rationale behind the visualisations' design, and then refining the prototype through gathering feedback from representative domain experts who are also current users of the software. Through a number of example scenarios, we show that the visualisation can support the identified tasks and aid analysts in discovering and understanding potentially risky insider activity within a large user base.
} 

\CCScatlist{
  \CCScatTwelve{Human-centered computing}{Visu\-al\-iza\-tion}{Visu\-al\-iza\-tion application domains}{Visual Analytics};
  \CCScatTwelve{Security and privacy}{Human and societal aspects of security and privacy}{Usability in security and privacy}{}
}

\begin{document}


\firstsection{Introduction}

\maketitle
\thispagestyle{fancy}

Last year’s Twitter attack \cite{Sec81} showed the potential severity of insider threat when cyber-criminals coerced insiders to alter user accounts to their financial benefit. The sums stolen were relatively small (around \$120,000 in Bitcoin), but temporarily knocked nearly \$1Bn off Twitter's market value, and caused embarrassment and inconvenience to the company and the figures whose accounts were compromised. This type of attack is on the rise and estimated to currently be the biggest single source of loss to cyber-criminality, coupled with ever-increasing penalties for the after-effects of losing or leaking personal information. To this end it is necessary for organisations to be able to better monitor and control for potential insider threats, and we present work that utilises the strengths of interactive data visualisation to help accomplish this goal.

\subsection{Insider Threat}
Historically, the biggest challenges in the cybersecurity domain were external in nature such as viruses in downloaded files or portable media, denial of service attacks on networks, or hacks that took advantage of hardware or software flaws. However, now an ever increasing share of threats come from insider sources, actions by individuals within an organisation that either purposely or unwittingly compromise digital assets. A recent market report by Gartner \cite{Sec167} indicates that an average insider threat incident costs over \$11M for the company impacted, a cost that has risen 31\% in just 2 years, and the number of such incidents has grown by an even greater amount, 47\%, over the same period.

The first line of defence against insider threat is acknowledged to be user education e.g. don’t accept files or click on URLs from unverified sources, and use strong authentication methods on accounts and portable equipment etc, so that third parties cannot obtain credentials and misuse legitimate accounts. This, however, only covers the risk of negligent and compromised exposure but not malicious or coerced behaviour. This differentiation of ‘insider type’ is a fundamental classifier in insider threat taxonomies, as seen in the majority of surveys on the subject \cite{Sec33,Sec38,Sec45,Sec63} - though in the earliest work insider threat was seen solely as the act of purposeful individuals \cite{Sec102}.

Further differentiation on malicious insiders can be made by ascribing motivation to financial, personal or political ends – people who are leaving an organisation (or who have already left without having their credentials revoked) are seen as a particular risk, as acknowledged by Homoliak et al. \cite{Sec33} in their 5W1H framework. Here, as in most frameworks regarding insider threat, it is the user, their psychology and their actions which are the focus of attention rather than the technology - all the surveys referenced here with the exception of Salem at al. \cite{Sec45} consider psychological drivers for at least malicious behaviour, and for some work \cite{Sec103,Sec104} it is the primary focus. Recently, work \cite{Sec166} has also looked at various methods of ingraining mitigation strategies into corporate culture, such as double-checking work and pair programming, but this obviously comes at a larger cost in human resources.

Technical approaches as used in UEBA (User and Entity Behaviour Analytics) systems are a vital, complementary element in tackling this risk. The two main technical methods used for detecting insider threats are signature/misuse and anomaly-based detection. The first uses rules that are triggered when actions meet certain thresholds, e.g. they occur outside normal working hours or access particularly sensitive resources. An immediate advance on this is to have rules triggered on given sequences of actions. Almost from the technique’s conception, it has been acknowledged to produce overwhelming numbers of false positives \cite{Sec19}, thus looking for the truly dangerous event is a needle in a haystack operation. The second method works on the assumption that potentially dangerous insider activity is indicated by actions outside the norm for a user or their role or account \cite{Sec100} - though malicious insiders will try to cover their tracks. Knowing this, the aim is then to efficiently find that activity that is outside the norm, by comparing actual to ‘normal’ activity either through rules or structures of increasing complexity or, latterly, data mining \cite{Sec3}, machine learning and deep learning techniques \cite{Sec101}. Still, the vast amount of this anomalous activity also turns out to be benign.

As both patterns of violation and one-off violations can be indicative of real insider threats and, when appropriately designed, one of the great strengths of data visualisation is to reveal patterns and outliers in data, then visualisation becomes a natural technique to explore and analyse insider threat. While academic research prototypes have investigated advanced visualisations for insider threat datasets, commercial software has been slower to adopt requirements-based visualisations beyond the standard dashboard of pie and bar charts. This paper thus describes the work involved in introducing visualisation techniques into FortiInsight \cite{Sec168} - a commercial software application that allows cybersecurity analysts to monitor and detect potentially dangerous user actions.

\section{Background}

Specific work on insider threat visualisation has progressed from early work by Colombe and Stephens \cite{Sec19}, through applications of data visualisation for specific insider threat detection methods \cite{Sec49,Sec30}, examination of the further challenges \cite{Sec87}, and onto full-bodied visual analytic approaches that take fresh approaches to the data analysis as well as the visual representations \cite{Sec50,Sec51,Sec15}.
 
Data visualisations are obviously dependent on the type of data, which in turn is decided by the type of insider threat detection employed and the raw data it in turn works upon. Nance and Marty \cite{Sec49} used role-based data to build a bipartite graph composed of users and resources, and by visualising the graph can reveal those combinations of user and resource interaction that are rare and/or unexpected for a given user's role. The graph visualisation is a natural aggregation such that each edge can represent multiple events and scales much more elegantly than displaying individual events. Haim et al. \cite{Sec30} displayed a typical dashboard of ‘top tens’ of users for current and cumulatively risky behaviour – this both aggregates (events rolled up by user) and filters the data (top 10) such that the visualisation is not overwhelmed. Here, risk is a quantity calculated from the user’s triggering of numerically rated rules, the underlying engine being signature based. 

Visual analytic solutions aim to both improve the analysis and visualisation of data, with the ambition being that together they can be more than the sum of their parts. Recent work has looked at general event analytics and event sequences in particular as a worthwhile avenue \cite{Sec57,Sec55,Sec10}, with Adnan et al. \cite{Sec3} exploring methods for identifying both sequences and unordered sets of events. The focus on event sequences both in the analysis and visualisation phases indicates that the temporal ordering of events is seen as important – accessing a sensitive file and then removing a pen drive is not as noteworthy as the other way round. Arendt el al. \cite{Sec5} built a visualisation over the CERT insider threat dataset \cite{Sec169} that displayed compressed (and hence aggregated) event sequences, which themselves emanated from a guided machine learning analysis of event data. Their visualisation combined glyphs and co-ordinated multiple views with sorting, filtering, and searching capabilities to help analyse the compressed sequences.

The most ambitious body of work relevant to insider threat visualisation is associated with the EU DiSiem Project \cite{Sec3,Sec50,Sec51,Sec15,Sec14,Sec52}. Here, researchers explored novel avenues of user behaviour analysis such as sequence mining with subsequent clustering \cite{Sec51}, topic modelling of event sequences \cite{Sec14}, and topic modelling fused with hierarchical user profiles (user roles/groups) \cite{Sec50}. Each has an associated and detailed visual interface with multiple views including overviews and timelines, and the ability to drill-down/aggregate, filter, search and re-analyse data. All are designed to fulfil tasks elicited first-hand from cybersecurity professionals. Here the data was not insider threat data as such, but events generated on a single login and security server by whoever accessed it. 

Further to this research, it is notable that other commercial insider threat products tend to be unadventurous in the types of visualisation employed - pie charts and bar charts abound in a typical dashboard. This is not automatically a bad thing, keeping it simple is a sound strategy but these visualisations tend also to focus on high-level patterns and revert to text-hunting for details. A few do employ more advanced visualisations, and while some of these have a tendency towards being marketing visuals (i.e. 3D networks) and chart junk, Securonix \cite{Sec171} and Splunk \cite{Sec170} in particular have both shown examples of techniques such as Sankey Charts, treemaps and radial charts within their products. Both though tend towards being fully-fledged SIEM products and it is hard to tell how and why they arrived at using these particular visualisations. 

It must be noted that the visualisation work surveyed here focuses almost 100\% on event data emanating from user actions - the other strands such as social mitigation or education aren't integrated. Haim et al. \cite{Sec30} do mention sharing information with other organisational departments such as Human Resources (with the assumption being to increase a user’s risk if they are under disciplinary action or is due to leave shortly), but little or no mention is ever made of a user’s psychological state, primarily because that is much harder to capture and evaluate, both ethically and technically.
 
In summary, the documented work closest in nature to ours is that of the DiSiem Project, but there are several important distinctions. Firstly, we are not re-analysing the data but interpreting it as is from a source database, where events have already been judged against criteria to determine their elevation to alert status or not, and we are focusing on the policy alerts only. Secondly, the dataset we use derives from not just one server but from many thousands of endpoints - roughly the same number as we have users. Thirdly, we have a larger array of possible event combinations compared to DiSiem's higher-level vocabulary of roughly 300 possible operations on a login server - ignoring users and timestamps, each of our events is still a combination of any resource (file or drive), any particular application, any of thousands of endpoints, and a CRUD operation. This differentiation also holds with the work by Arendt et al. \cite{Sec5}, who acknowledge that in practice real datasets are likely to contain many more unique actions than the CERT dataset they employed. In practical terms, this means our data is much less amenable to the sequence mining those approaches utilise.

\section{Design}

\subsection{Glossary}
To aid comprehension, in \autoref{tab:glossary} we give terminology which we aim to use consistently throughout the paper.

\begin{table}[h]
    \centering
    \caption{Glossary}
    \label{tab:glossary}
    \begin{tabular}{ |p{1.4cm}|p{5.7cm}| } 
     \hline
     \textbf{Term} & \textbf{Meaning} \\ 
     \hline
     User & Or more correctly, a user account. An actual human can have access to more than one user account, and more than one person could access the same user account (a possible insider threat!). Typically though it is a one-to-one relationship between accounts and users. \\
      \hline
     Event & An action performed by a \textbf{user} such as opening or copying a file, or inserting a USB stick that is registered by the software. An event generally contains the endpoint, user, time, application, resource and CRUD action - i.e. user account Bill on endpoint WORKPC1 used Word to create file top.docx at 11:03. \\ 
      \hline
     Policy & A pre-defined rule that an \textbf{event} can be judged against. May contain multiple (AND) and disjoint (OR) clauses.  \\
      \hline
     Anomalous & An \textbf{event} that is markedly different to normal given a \textbf{user's} past activity as assessed by an AI component. \\
      \hline
     Alert & An \textbf{event} that is \textbf{anomalous} or satisfies a \textbf{policy} definition is upgraded to alert status. Multiple closely timed \textbf{events} that meet this threshold may be aggregated into one alert. \\ 
     \hline
     Endpoint & A device on which \textbf{event} capturing software is installed. Typically a PC, but also can encompass servers and printers. \textbf{Users} and endpoints are generally one-to-one but not always. \\
     \hline
     Analyst & A person who analyses and explores the \textbf{event} and \textbf{alert} data sets. Analyst is used in this text to distinguish this role against the \textbf{user} who is the person whose actions are being monitored. \\
     \hline
    \end{tabular}
\end{table}

\subsection {Methodology}

We proposed to develop the visualisation using a tried and tested methodology that had worked previously. Firstly, collate the analysts' tasks/requirements, understand the data and get a feel for how the analysts worked. From this we could design specific visualisations to accomplish specific tasks and confidently expect to construct an interface that fulfilled the analysts' needs. Iterative feedback and interaction with real analysts would keep the work on track and stop the development moving off at a tangent to real needs.

\subsection {Requirements}

Collecting analyst tasks that the visualisation should fulfil was initially planned as a direct user-facing exercise, but this had to be dropped in the face of Covid restrictions. We already faced the problem of access described by Grudin \cite{Sec1} who decades ago established the difficulties of attempting to engage actual end-users from within a commercial setting where well-intended barriers are established between development and sales, and the purchasing customers are often not the same people as end-users, or even in different organisations if sold via software resellers.

Fortunately, cybersecurity literature exists that elucidates tasks and Kerracher and Kennedy \cite{Sec37} show that it is acceptable to derive tasks from literature if the domain terminology is understood and a similar problem is being tackled (in this case, detecting / assessing insider threat). With that in mind, we reviewed papers that covered cybersecurity visualisation, insider threat analysis and event analytics that specified appropriate tasks we considered amenable to visualisation \cite{Sec87,Sec50,Sec51,Sec5,Sec30,Sec23,Sec84,Sec56,Sec68,Sec86,Sec105} to collate a list of candidate tasks. This gave us over 80 tasks, many of which were going to obviously overlap, eg. Legg \cite{Sec87} had 'Find user performing activity at unusual time', and Arendt et al \cite{Sec5} had 'Do events occur outside normal working hours'. Often the difference was merely terminology - 'See alerts that occur around the same time as a specific alert' v. 'Study antecedents or sequelae of an event of interest'. Therefore, we used Brehmer and Munzner's \cite{Sec7} abstract visualisation task typology to classify them by what each task did (the why typology) to aid finding replica tasks, and then merged these similar tasks into a single item described with a consistent vocabulary, which then placed them into natural categories such as View, Compare, Find, Generate and Summarise. At the end of this process we had roughly 20 visualisation-amenable tasks along with a few more general usability-oriented tasks such as history, report generation and extracting data for further processing. 

To explore which of these were considered relevant by analysts we made contact with the cybersecurity group of a large customer of the product and obtain permission to send them a short survey and an interview with the group leader. The survey simply asked respondents to rank the tasks on a scale of 1-5 for perceived usefulness and, while there weren't enough responses (eight) for any rigorous statistical validation, some patterns did emerge. Firstly, it was noted that the comparison tasks were ranked bottom across all the responses. The group leader explained this: the bulk of their work was watching for users triggering their policy settings on a day-to-day basis and they were interested in which users did this, when, and what they were doing. It was interesting to see who was causing the most alerts as that is obviously an indicator of where more risk can occur, but this was seen in an absolute rather than comparative sense. Only in limited circumstances would they be interested in specifically comparing user behaviour, though seeing if a user triggered more alerts than normal was rated slightly higher.

The interviewee also stated that they valued the policy alerts over the AI alerts. The policies had been set up at least partly by the analysts themselves so they understood why and what triggered them, whereas the AI alerts appeared as a black box - it was difficult to understand why the AI engine was flagging certain events as alerts and not others.

Of the non-comparison tasks, it appeared that the View and Find categories were held in highest regard (see \autoref{tab:tasks}), and none apart from "View role-orientated or task-orientated visualisations" gained less than an average score of 4.0 or were outside the top half of the rankings. Again though it must be stated that this was a very small sample. Extracting data tasks were also rated in the top half, as the analysts explained that after potentially risky alerts / users / activity were identified and explored the next steps tended to require passing on this data to other tools or people including Human Resources. This however is an operation that the existing software can do already. Of these we decided to concentrate on the View and Find tasks as they were particularly amenable to visualisation and we label them as T1-8.

\begin{table}[h]
    \centering
    \caption{Top rated tasks extracted from previous literature}
    \label{tab:tasks}
    \begin{tabular}{ |p{0.6cm}|p{6.0cm}|p{0.3cm}| } 
     \hline
    \textbf{Score} & \textbf{Task} & \textbf{ID} \\
    \hline
    4.63 & View details of specific user, event or alert & T1 \\
    \hline
    4.63 & Find user performing anomalous activity & T5 \\
    \hline
    4.57 & Extract a specific user’s typical activity & \\
    \hline
    4.50 & View events/alerts ordered by selected attribute (time, type, user etc) & T2 \\
    \hline
    4.50 & Understand extracted anomalous events & \\
    \hline
    4.50 & Find events of interest (by time, type, user etc) & T6 \\
    \hline
    4.38 & View relationships between users / events & T3 \\
    \hline
    4.33 & Understand extracted frequency of events & \\
    \hline
    4.25 & Find proximal events to a specific event for the same user & T7 \\
    \hline
    4.25 & View different data facets (users, events, alerts, locations etc) & T4 \\
    \hline
    4.25 & Combine different data facets (users, events, alerts, locations etc) & \\
    \hline
    4.25 & Find deviations from required/typical pattern of activity & T8 \\
    \hline
    4.25 & Summarise a specific user’s current or past activity & \\
    \hline
    \end{tabular}
\end{table}

\subsection {Dataset}
The dataset in this paper was used with permission from an existing product customer with a userbase of over 15,000 individuals. It comprises of an ElasticSearch \cite{Sec174} database with nearly 900,000 policy alerts collected through a period of eventually over 2 years (27 months). Each alert contained the array of raw events that had triggered the policy, along with the time it was triggered, and policy-specific information such as the name and id of the policy and associated severity (a pre-determined measure of how dangerous the analysts considered triggering the policy to potentially be). As this is a confidential data set it unfortunately cannot be openly shared.

\begin{table}[hb]
    \centering
    \caption{Common event and alert attributes}
    \label{tab:event}
    \begin{tabular}{ |p{2cm}|p{5.1cm}| } 
    \hline
    \multicolumn{2}{|l|}{\textbf{Event Attributes}} \\
    \hline
    Who & User \\
    \hline
    What & Application, Resource \& Activity \\
    \hline
    When & Start \& End Time \\
    \hline
    Where & Endpoint \\
    \hline
    \multicolumn{2}{|l|}{\textbf{Alert Attributes}} \\
    \hline
    What & One or more Events\\
    \hline
    When & Alert Time\\
    \hline
    Why & Policy (Policy Breach) or Tag \& Confidence (AI Detection) \\
    \hline
    \end{tabular}
\end{table}

\subsubsection{Data Profile}
Initial analysis of the alert database with Elastic's interactive Kibana \cite{Sec172} GUI revealed both data quality issues and some global features of the data:

\paragraph{Data Quality}Firstly, there were large spikes in alert totals in the first weeks of the dataset and around October 2020 - one week alone accounted for 140,000 (20\%) of all alerts. Discussions with an analyst revealed this to be caused by issues setting up the policies initially and later with introducing a new policy.

Secondly, a few of the alerts contained over a thousand events each, again all occurring within the first few weeks of the data set, which again was put down to the same teething difficulties - most were caused by long-playing video files which caused events to repeatedly fire as the user continued to watch it and were then bundled up into the same alert. This had been previously noticed by the analysts, and after this a hard limit of 100 events per alert had been enforced.

Later, and using the prototype itself, we found one particular user account had caused over 100,000 alerts just by itself. This turned out to be a pseudo account that was firing the same alert repeatedly on the same application, starting after a particular date.

It was agreed that the problematic date ranges and the pseudo-account could be excluded from the visualisation using a set of appropriate clauses that were applied to all queries. This eliminated both the extreme spikes in weekly activity and the overloaded alerts.

\paragraph{Data Shape} Following this cleaning we re-analysed the data, and saw that over 13,500 users had caused at least one alert over the course of the dataset, but there was an obvious long tail distribution. Even within just the top 100 users, the top user had triggered 30 times more alerts (14,000 vs. 500) than the 100th placed user. The general pattern of a roughly logarithmic distribution of events per alert was also observed, even after excising alerts with 1000+ events, and most alerts (66\%) contained only a single event. Also, when multiple events were bundled into the same alert they always concerned the same user, endpoint and application, while actions and resources could vary. It was also noted that most of the alert/event properties were categorical in nature, only a few such as the temporal attributes and severity/confidence scores were naturally continuous or numerical.

\subsection {Interface \& Visual Design}
\begin{figure*}[h]
  \centering
    \begin{tikzpicture}
        \node[anchor=south west,inner sep=0] at (0,0) {\fbox{\includegraphics[width=.98\linewidth]{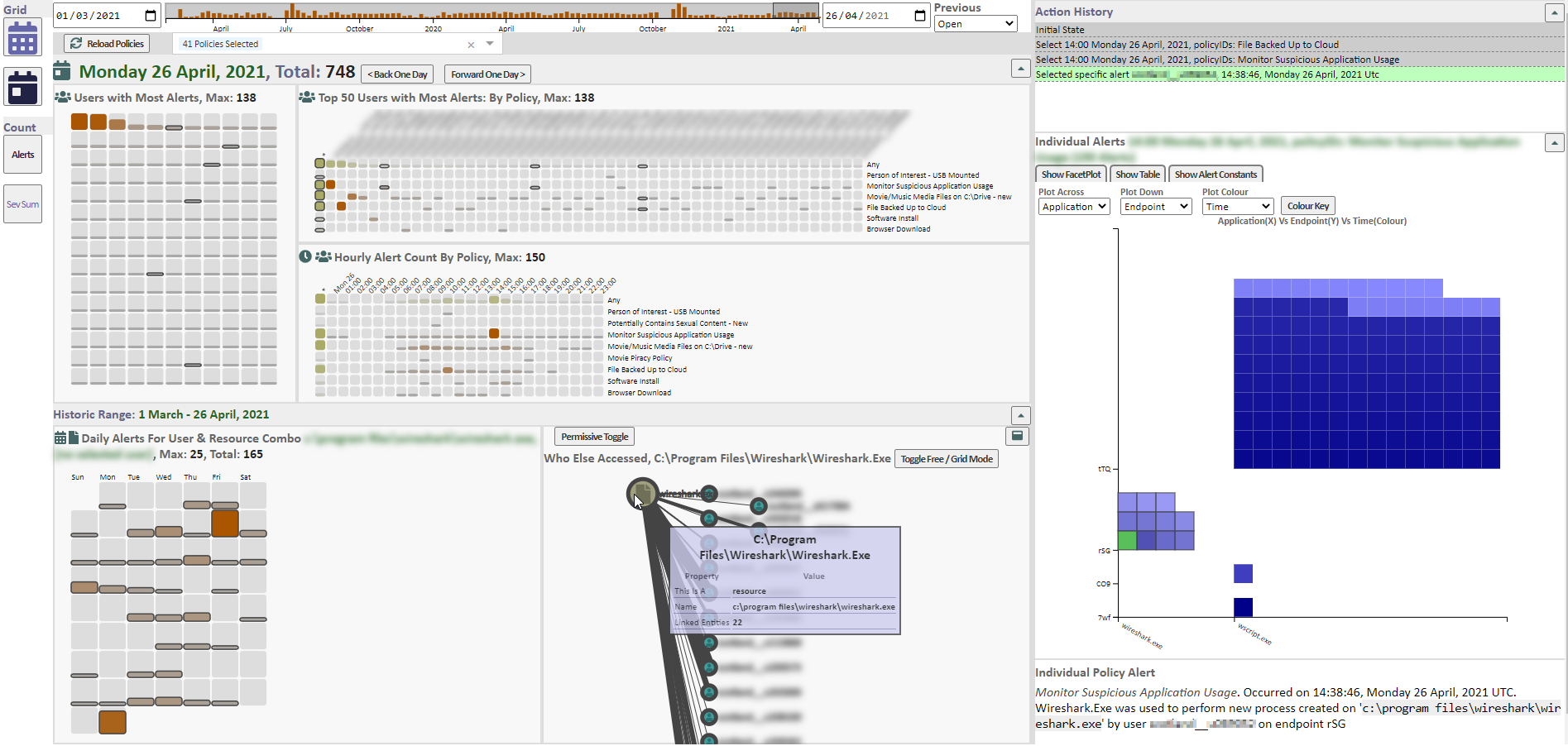}}};

        \node[text width=0.2cm, fill=white, opacity=.7, text opacity=1] at (1.6,8.25) {\textbf{A}};
        \draw[red,thick,rounded corners] (1.8,8.5) rectangle (9.5,8);
        
        \node[text width=0.2cm, fill=white, opacity=.7, text opacity=1] at (0.4,5.45) {\textbf{B}};
        \draw[red,thick,rounded corners] (0.6,7.2) rectangle (11.4,3.7);
        
        \node[text width=0.2cm, fill=white, opacity=.7, text opacity=1] at (11.7,3.75) {\textbf{C}};
        \draw[red,thick,rounded corners] (11.5,6.5) rectangle (17.5,1);
        
        \node[text width=0.2cm, fill=white, opacity=.7, text opacity=1] at (0.4,1.9) {\textbf{D}};
        \draw[red,thick,rounded corners] (0.6,3.6) rectangle (11.4,0.2);
    \end{tikzpicture}
    \caption{The FortiInsight visualisation explorer. Some details have been redacted for customer confidentiality. The top bar (A) shows a weekly aggregate of alerts, the three grids in the top left (B) show varying aggregations of the current day's alert activity, the right-hand side (C) shows individual alerts from a chosen aggregation element and the bottom left (D) shows a grid and network view of past user and resource interactions.}
    \label{fig:annotated}
\end{figure*}

The final interface for viewing the policy alerts has a number of related sub-views as shown in \autoref{fig:annotated}, in a division that roughly fulfils operations according to Shneiderman's mantra \cite{RN142}. We deliberately tried to keep the visual design and properties simple and not introduce visualisations simply for the wow factor, as the analysts were used to the product's current interface of bar and line graphs and trying to keep some familiarity was viewed as helpful.

\paragraph{Overview - High Level}
The top view (\autoref{fig:annotated}A) shows a histogram of the entire 2+ year dataset, aggregated by week, which acts as an \textit{overview}. A user-defined brush within this histogram acts as a control to set a sub-range of dates for the rest of the interface, effectively acting as a \textit{zoom} on the \textit{overview}. This view itself indicates particular weeks that are outliers in terms of number of alerts (T8) - on one of the revised data snapshots, it helped reveal a policy / user combination that was generating 90\% of all alerts after a given date.


\paragraph{Overview - Aggregated Views}
Below this top view are sub-views showing data that falls within the zoomed date range (\autoref{fig:annotated}B). At this scale the number of alerts is still large so aggregated counts of alerts are represented rather then individual alerts. Grid views comprise of a number of squares - each square representing a unit of aggregation, the unit depending on the particular make up of the grid, with a particular ordering. Within each grid a redundant visual encoding of bar size and colour saturation is used, both increasing with the number of alerts. Each grid can have restrictions in that it may be limited to a certain time range or just one day, or report on alerts for just one user rather than all users. The current combinations that are employed are detailed in \autoref{tab:grids}.

\begin{table}[htb]
    \centering
    \caption{Grid view combinations}
    \label{tab:grids}
    \newcolumntype{P}[1]{>{\raggedright\arraybackslash}p{#1}}
    \setlength{\tabcolsep}{0.5em}
    \begin{tabular}{ |P{1.6cm}|P{1.4cm}|P{2.0cm}|P{1.9cm}| } 
    \hline
    \textbf{View Name} & \textbf{Agg. Unit} & \textbf{Order} & \textbf {Filter} \\
    \hline
    Calendar & Day & Day & Day Range\\
    \hline
    Daily Top Users & User & Alert Count & Single Day\\
    \hline
    Historic Top Users & User + Day & Alert Count & Day Range\\
    \hline
    Single User Calendar & Day & Day & Day Range \& Single User\\
    \hline
    Daily Top Users by Policy & User + Policy & Alert Count \& Policy Severity & Single Day\\
    \hline
    24 Hours by Policy & Hour + Policy & Alert Count \& Policy Severity & Single Day\\
    \hline
    Targeted Calendar & Day & Day & Day Range, Selected Users \& Resources\\
    \hline
    \end{tabular}
\end{table}

\begin{figure}[h]
 \centering 
 \includegraphics[width=\columnwidth]{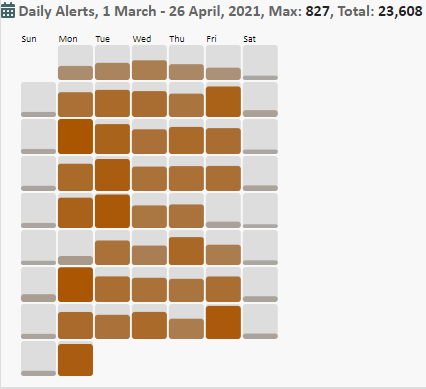}
 \caption{Calendar grid, showing alerts per day for the selected date range.}
 \label{fig:calendar}
\end{figure}

The design can be considered simply as a bar chart that wraps columns rather than continuing as a single horizontal line - this allows weekly or day of the week patterns to be observed when set up as a calendar-like view, and a space saving device when showing a grid ordered by alert count. The 'Daily Top Users by Policy' and '24 Hours by Policy' combinations show the data in a manner analogous to a stacked bar chart, but 'exploded' such that each row/series can be judged by a baseline rather than the trickier comparison of trying to compare sub-parts of the stack by length \cite{RN1195}. In this manner the analysts can see the worst days, the worst users (T2, T5 and T8) and particular combinations thereof.

\paragraph{Details-on-demand - Facetplot}
Selecting a grid square will then retrieve the individual alerts that aggregation represents and show them in the facet plot (\autoref{fig:annotated}C), which acts as a detail view. This is essentially a categorical scatterplot where two selected properties from the alerts and their bundled events can be used to plot the alerts on the X and Y axis in groups. This approach was chosen as the majority of data attributes were categorical and a traditional scatterplot would have displayed a lot of overlapping elements at a few discrete points.  We did consider the use of N-dimensional visualisation techniques such as parallel coordinates \cite{RN340} and scatterplot matrices, but the mostly categorical nature of the data would have had led to the same drawbacks as with the single scatterplot. A third property can be selected to colour the alerts with a graduated saturation of a single hue (blue) for continuous properties and a desaturated colour scale for categorical properties - this last was chosen to avoid a bright red option, because an early internal demonstration with another scale showed people attached a 'danger' meaning to anything marked in that colour. An attempt at utilising the facetplot as a UMAP-based \cite{Sec175} Multi-Dimensional Scaled plot was rebuffed by the analysts (see 4.1). 

This view fulfils T1, T4, T6 and T7 - the analyst can segment the data by different attributes (T4) and then find alerts of interest by those attributes (T6). Finding proximal alerts is simply a matter of seeing which other alerts are grouped with, or grouped nearby to a given alert (T7). A tooltip can reveal exact details of an alert, and a small text panel underneath shows details for a selected alert (T1).


\paragraph{Relate - Node-Link View}
Selecting an individual alert will also send the alert details to a node-link view (\autoref{fig:annotated}D), which reveals which other users were involved in alerts with the same resource, and in turn which other resources were involved in those alerts. As this view was specifically meant to show relations, a node-link view was a natural fit to this information.

Nodes are distinguished by type using a redundant combination of colour and icons and sized on a logarithmic scale by the number of alerts the user or resource is involved in. Edges are drawn as simple lines with thickness correlating to the number of alerts between a user and resource. Hovering over a node will highlight related edges and nodes, and hovering over an edge will highlight related nodes. Direct selection of a node will populate a final grid view with the daily history of alerts for that user or resource. If an edge is selected only alerts involving both the user and resource are plotted in the grid view. This grid view in turn can be used to select aggregations to send back to the facet plot. Trying to find a particular node in a general node-link view is often a laborious task, so there is also the option to regiment the layout - here the nodes are laid out in rows and columns by type in alphabetical order, as in \autoref{fig:annotated}D.

The view satisfies T3, as analysts can now see the relationships between users and resources - the main attribute of further interest when dealing with alerts (see Section \ref{third focus group} for the discussion on why resource acquired this status, rather than endpoint or application).

\paragraph{History - History View}
A history of actions within the interface is recorded and shown as a list. Selecting an entry within the list will return the visualisation to the state those settings represent. Branched histories are also supported: items from which divergent exploration paths are started are shown as tabbed selections - selecting a tab reveals the list of actions continuing on from that point. We also designed the application as a single page app - panels could be hidden or expanded if needed for more detail - and between this and the history view, returning to a previous point in exploration became much easier than in the products' existing UI.

\paragraph{Filter}
While the brushed range in the overview acts as a zoom, it can also be considered a temporal filter. The ability to select subsets of alerts from the aggregate views to forward onto the facet plot are also implicit filters. Further to this, it is also possible to filter on individual policies if only some are currently of interest, and to select one particular user to focus on. In this sense, these filters help satisfy T6 as they enable the analyst to focus on alerts of interest by user, time or policy.

\subsection {Implementation}
The visualisation was built as a single-page web application using D\textsuperscript{3}\cite{RN1099} and Angular, with D\textsuperscript{3} being used within individual sub-views to display the data and Angular performing the macro-level communications and control between the sub-views. The web application obtained data via a bespoke API layer which formed a shim over an ElasticSearch database. 

\section {Analyst Feedback}
We have been in touch regularly with the customer's analysts to run remote focus groups which have proved vital in shaping the development of the visualisation. Such domain experts are the gold standard for evaluation: willing cybersecurity students could be employed but lack the experience of employed analysts, and the analysts have the further motivation that they are being presented with something that could well form part of their day-to-day work in the future rather than just an interesting diversion. Using Mazza and Berrè's \cite{RN897} focus group template as a guide we gave remote demonstrations and then asked pertinent questions about how they felt the visualisation aspects worked and whether it was doing what they expected. General feedback was also solicited, which turned out to be a rich source of information. Each group generally had up to three analysts, plus a researcher leading the group, and occasionally a developer who helped with context and terminology.

\subsection{First Focus Group}
The first focus group used a dataset of 130,000 AI alerts covering a single month - these are similar to policy alerts with, instead of a policy name and severity, a tag string indicating what aspect of the event caused the AI to trigger an alert, and a confidence rating as to how sure it was that the event is anomalous. The focus group reported that using date and user as the initial units of investigation was a correct thing to do - it is after all users that are the ultimate source of insider threat, not applications or CRUD operations. They also noted that temporal patterns in the grid widgets were easier to spot than in the regular product interface.

\medskip
\textit{
“It’s useful to see a person has triggered this on multiple occasions, across different days, because that could lead to patterns – in aggregate might not see user has been up to stuff on multiple different days and what the pattern is”
}

\medskip
\textit{
“that’s what the system’s meant to do, absolutely should be the focus of it – you could filter on just certain types of event too on top of users”
} (Focusing on the individual)

\medskip
They also stated their wish to quickly select a specific individual which reflected task T1.

\medskip
\textit{
“We could be investigating a particular user and the ability to enter that person’s id and see on a page what has that person doing for the past 24 hours / 7 days / 30 days example and tailored towards that specific user, rather than the overall picture of our whole estate”
}

\medskip
\textit{
“What *** suggested, just drill down to one user and see quickly have they written anything to usb and then deleted it, or used a usb in general, things like that if you suspect a specific user”
}

\medskip

However it was at this point we discovered their underlying concerns about the AI alerts - they were not sure exactly why they were being fired and the overwhelming number of them was also a concern. 130,000 alerts over a single month was roughly 5-6,000 per working day, and shared between 3-4 analysts was 1,500-2,000 each per day - it was easy to see that for each analyst investigating 200 alerts an hour, or 3 a minute, wasn't feasible. Even if the confidence rating was used as a filter (there were many more low-ranked confidence than high-ranked confidence alerts) the debate returned to whether they were the truly worrying alerts, or just the most anomalous. It was at this point that they stated their preference for policy alerts, and the focus groups after this used the policy alert data set previously described.

\medskip
\textit {
“The policy stuff is good because they’re absolutes and if you’ve created that criteria you obviously want to know about that specific action happening, whereas the AI’s a bit more black-box kind of thing – we don’t quite know what’s caused that to happen.”
}

\medskip
\textit {
“Yes, because that’s what (other colleagues) spend their time investigating on a daily basis. The AI is more of a black box.”
}

\medskip

This isn't due to any peculiar kind of reticence on their part - A Gartner Market Report from 2018 \cite{Sec176} states that \textit{“Contrary to many vendor claims, UEBA solutions are not 'set and forget' tools that can be up and running in days. Gartner clients report that it takes three to six months to get a UEBA initiative off the ground and tuned to deliver on the use cases for which they were deployed. This jumps to 18 months for more complex insider-threat use cases in large enterprises.”}

Interestingly, when demonstrating the facetplot, we were aware that trying to find patterns by trialling different combinations of alert properties to attach to the axes could be laborious and time-consuming, so we explored using multi-dimensional reduction (MDR) to produce a two-dimensional view of alerts over the full dataset and property range. UMAP (Uniform Manifold Approximation \& Projection) \cite{Sec175} is a recent MDR technique which has proven to be computationally faster than other techniques \cite{Sec177} and was thus chosen to demonstrate the process. The UMAP algorithm would be run over the alerts' properties and two extra numerical fields, UmapX and UmapY, were returned which would be used to arrange the facetplot visualisation. However, the feedback from the analysts was less than positive as, like the AI alerts, it didn't explain why certain clusters appeared, the result was just given as bald fact. Setting the colour variable could show how individual properties were distributed but also still didn't explain \textit{why}. Their preference instead was to recommend setting up default choices for the axes using the properties they considered the most discriminative.

\medskip
\textit {
“In what way are they similar? I can’t see a lot of use for this, without a lot more explanation of why clustering happens. Unless we were looking for something very specific – like we knew an alert for ransomware existed, what was similar – but very specific circumstances”
}

\medskip
\textit {
“The tagstring right away, I want to see what caused these spikes, I would leave the configurability in place though people might want to use different options depending on situation”
}
\medskip

\subsection{Second Focus Group}
Overall, the opinion from the first focus group was that the visualisation was a definite improvement over standard commercial UIs for insider threat products. Taking their feedback into account, we re-designed the interface to refocus on the policy alert dataset and to have a mode that started on a particular day to fit with their dominant method of working - triaging new alerts on a daily basis. This was demonstrated at a second focus group which concentrated on how useful the visualisation could be for the analysts' day-to-day work.

We demonstrated the grid view where the top 50 users were laid out in columns and the policies in rows, and explained this showed not only which users or policies were the most troublesome but also the combinations thereof. The analysts' opinion was that if one policy is dominant their attitude is that the policy is probably over-sensitive rather than hundreds of employees are up to no good. For example, a policy relating to playing mp3/mp4 files was the cause of most alerts on many days, but this was explained as people listening to their own music while working - especially as the bulk of the customer's office-based employees were working from home due to covid guidelines. The analysts would probably then tighten the terms of that policy to reduce the number of alerts it caused.

\medskip
\textit {
“For that policy, it shows it maybe triggers too much, everyone is working from home and playing music or something”
}
\medskip

On a visualisation specific note, one analyst noted they would expect the policies to be ordered by their assigned severity, rather than the count of alerts they triggered. This is because they stated they were more interested in policies that fired a few times but were ranked as having a high severity, as opposed to a low severity policy that produced more alerts.

The analysts said the historical view of the entire dataset was interesting, but when they went looking for historic policy data it would be with a specific target in mind, such as a user or resource, rather than for an overview of general patterns - and that specific target would be because of something revealed in the current day's data. In that sense, historical data would be useful if restricted to a specific user - or in a list of decreasing probabilities: a specific resource, endpoint or application.

\medskip
\textit{
“The useful ones for us would be user, endpoint, application, resource. Activity sometimes because it tells us when a USB is mounted – that’s what we use to create the policy. But I think those three, endpoint, user, resource are the things we use more.”
}
\medskip

At this point we had also introduced a small timeline for selected users showing when in a 24-hour period they had triggered alerts. However, the analysts stated that it was out-of-hours alerts by policy they were more interested in. For instance, one policy relating to sexual content was viewed as much more noteworthy if it triggered alerts outside normal working hours - within working hours there were some staff working on issues around sex education so some alerts could be explained at that time but less so at 3am in the morning. Therefore they said it would be more useful to see an hourly breakdown of alerts by policy. Essentially though, it was for the analyst to examine the details of the alert and decide if it was benign or required escalation.

By now, we had a better understanding of the threat-hunting aspect of the work: find a peculiarity on a particular day, and then go back and see if any of the entities involved - user, endpoint, resource - had anything interesting to show in their past activity.

\subsection{Third Focus Group}  \label{third focus group}
Taking the previous feedback we developed a node-link graph view showing which users accessed a selected endpoint, and then in turn which other endpoints they had accessed in their other alerts, and showed this as the main point of the third focus group. Here though, while the effort was appreciated the analysts commented having seen it in the flesh, they would prefer to see resources in such a view instead. Endpoints were mainly one-to-one with users and only endpoints which were servers would show accesses by multiple users. It was also explained that while some computers were hotdesked in offices, Covid guidelines meant people were sitting at home with laptops and not sharing PCs.

\medskip
\textit{
“I think it would be more for resources than endpoints, because usually if an endpoint is used by many people it is a shared computer or a server, usually most people have their own computers, so we would be more interested in seeing for resource, who’s accessing that resource or who’s doing something with that resource and not who’s using that endpoint. I mean, sometimes, maybe we would use this, but it’s more for resources – most of the laptops we have are personal”
}

\medskip
\textit{
“There are hotdesk machines, at the moment everyone has their own machine, I don’t know what will happen to them once we return to the office”
}
\medskip

The analysts also commented that they didn't particularly need to know who had accessed a resource or endpoint first. We had implemented a grid view of days against users that showed when users had accessed an endpoint and an alert had occurred, and while one analyst commented it did give more information it was in general a very sparse matrix with very small individual elements. This hinted that this view's space would be better allocated to other information.

\subsection{Fourth Focus Group}
At this iteration we had taken what the analysts said about seeing the relationships between resources and users into account, and set up the node-link graph to work with resources. This immediately made for a more interesting view of the data as users alerted on many more individual resources than endpoints, including the fact that each alert could involve multiple resources. Unlike endpoint IDs though, which are short consistent alphanumerical strings, resources were typically filepaths that caused problems with naive searching e.g. 'C:/Users/John/wscript.exe' wouldn't match 'C:/Users/Janet/wscript.exe'. We introduced a 'permissive' mode where the search query matched just on the filename segment of the filepath so users who had accessed a resource with the same filename would be included in the network. 

Secondly, we demonstrated a history function that showed the ability to detail and restore past states in the visualisation. Thirdly, and finally following up the final bits of feedback from the second focus group, we dropped the small user timelines and introduced a further grid view of hours of the day versus policies for alerts - this showed the 24-hour pattern of alerts by policy the analysts had said they'd be more interested in than by user.

The reaction to all was much more positive than in the previous meeting, especially for the user-resource graph and the history functionality. This was especially pleasing for the node-link graph, as we had wondered at the start of the project about how the analysts would respond to novel (to them) visualisations.

\medskip
\textit{
“I think the resource thing could be really useful, because basically I think where we told you last time that we would need – that many times someone asks us like, who has accessed these and we will have to like do the search and threat-hunting and it takes a long time, so I think seeing it like this - and also seeing the same person, like the same people, are using that resource, what other resources are they like using, maybe they’re related, I think that one’s useful, yeah”}
\medskip

The reaction to the history function was also markedly positive.

\medskip
\textit{
"Yes, it would be useful sometimes, because you start doing an investigation and maybe 20 minutes later you’re still doing the same but you don’t remember how you got there so I think it’s useful to know this led me to that other thing and that’s how I got here. I think that would be yeah…”
}
\medskip


Finally, the analysts said that the use of alert IDs in the interface wasn't useful, unsurprisingly as they are random hexadecimal strings. This was most obvious in the history view as alerts were described by id, so after the focus group they were quickly replaced by username and alert time. An option for a user to annotate important history items with their own descriptions may also be beneficial.

\section{Case Study}

To show how the visualisation as designed can support the tasks identified earlier, we present three example scenarios:  

\subsection{Daily Activity}

The most common scenario for our analysts would be to login in the morning and see what activity had occurred in the previous day. For this we constructed an interface that was setup with three grid sub-views: Daily Top Users, Daily Top Users By Policy and 24 Hours By Policy. The first two differed in that the second sub-divided user alerts by policy at the expense of needing more space and thus being able to show less users (50 compared to 300 for Daily Top Users).

\begin{figure}[h]
 \centering 
 \includegraphics[width=\columnwidth]{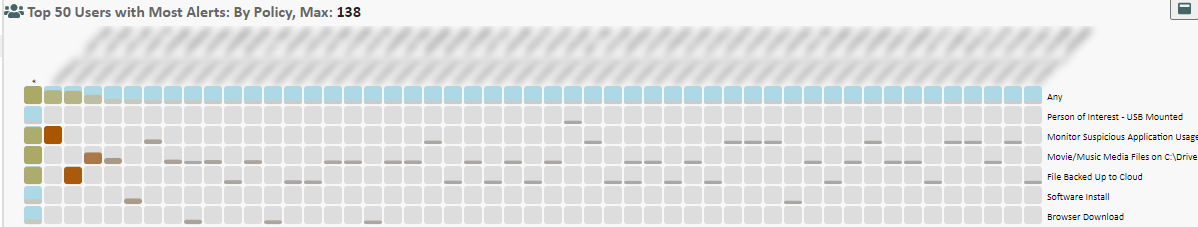}
 \caption{A grid view showing the top 50 alerting users (identifiers redacted) for a given day, sub-divided by policy. The most severe policies are at the top, and the top row and left column show totals.}
 \label{fig:usersByPolicy}
\end{figure}

Taking as an example the last day we have data for, the 26th April, we can see there are three users who immediately stick out as the leading causes of alerts (T5) in the Daily Top Users, shown top left. What's interesting is that the Daily Top Users by Policy (\autoref{fig:usersByPolicy}) shows these three users to have each triggered different types of policy - suspicious application usage, backing up to cloud, and using media files (T2) which is not revealed in the more compact view. 

Selecting the user with the suspicious application usage alerts sends those to the facetplot view. Here, the alert constants tab says these are all triggered by the application and resource wscript.exe on the same endpoint. Setting the facetplot to alert hour shows most of these alerts occurred between 2 and 3pm, a fact backed up by the Hourly Grid view. 

\begin{figure}[h]
 \centering
 \includegraphics[width=\columnwidth]{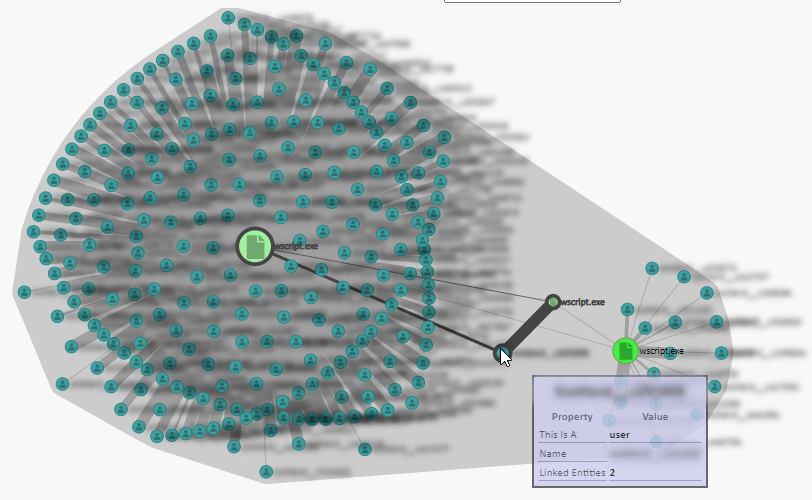}
 \caption{Node-link view showing the other users who have triggered alerts on versions of wscript.exe. One user sits between the obvious clusters having triggered alerts on two versions.}
 \label{fig:Scenario2b}
\end{figure}

There's not a lot more information to be gleaned from the facetplot, there's 100+ alerts involving wscript.exe that have occurred in a short period of time, so it would be useful to see if this is typical for this user or any user. Selecting one of the alerts populates a small textbox with details of that alert (T1) and sends that resource to the node-link view. Here a query is ran to show every user that has alerts involving that resource, which over the time period 1st March - 26th April involves nearly 200 users on wscript.exe (T3) (see \autoref{fig:Scenario2b}), the permissive toggle has been set on the filepath so system32 / win64 etc versions all report back). In summary, it shows that wscript.exe triggering an alert is not an uncommon activity.

\subsection{USB Stick Use}
One of the scenarios that the analysts said they were particularly interested in were low occurrence but high severity events.
In the last scenario, it can be seen in the 24 Hours by Policy View (\autoref{fig:teaser}) that while the number of alerts is dominated by two or three policies, the most severe policy category - someone of interest using a USB stick - triggers only three times, all for the same user.

\begin{figure}[h]
 \centering
 \includegraphics[width=4cm]{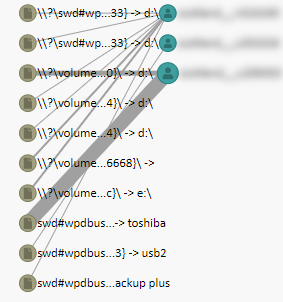}
 \caption{Users who have been involved in alerts with USB sticks with the same GUIDs.}
 \label{fig:Scenario3}
\end{figure}

Selecting this grid item produces just three alerts in the facetplot, so no slicing or dicing is possible here, but the tooltip reveals that the resource is now a descriptor containing USB volume GUIDs rather than a filepath. The analyst will now want to know how often this user has been accessing this USB stick so selects the first facetplot item: this reveals a simple node-link graph of the user connecting to three resource descriptors, one for each of the three alerts, containing between them three different USB GUIDs. This doesn't tell us anything extra, but by selecting the 'permissive toggle' we now tell the query that brings back the data for the node-link graph to match just on the GUIDs rather than the entire string. This now reveals more resource descriptors in the graph that show the same GUIDs involved in alerts for the original user (T7), but more interestingly, it introduces two new users who have been involved in alerts where the resource descriptor included at least one GUID involved in the original user's alerts (T3) - see \autoref{fig:Scenario3}. This is possibly evidence of a USB stick being passed around which is suspicious when shared drives are the accepted way of sharing data.

\subsection{Policy Sanity Checking}

Much of the noise in the dataset that we discovered in 3.4.1 was due to misfiring policies. For instance, we saw after one particular week that alerts shot up 5-fold and the Historic Top Users grid showed that the top 24-hour periods of alerts by user all concerned a system user account that was constantly triggering one policy (T5).

\begin{figure}[h]
 \centering
 \includegraphics[width=\columnwidth]{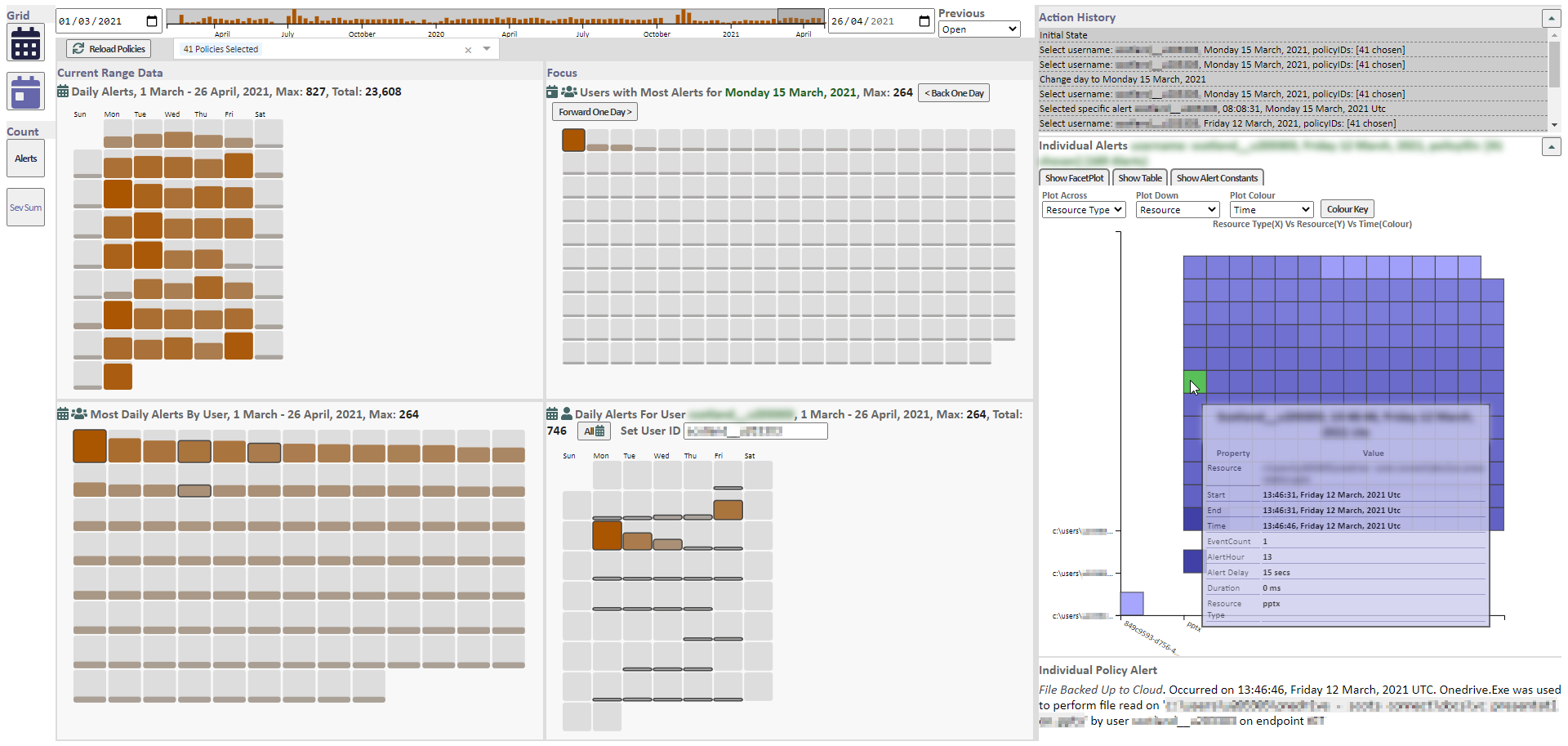}
 \caption{Historic Data View, showing the calendar view top left and the worst (most alerts) 24-hour periods for individual users bottom left. The worst user period has been selected and the concerned user's historic daily activity is shown bottom middle. Inspection in the facet plot shows these alerts were all triggered by one policy on one file.}
 \label{fig:Scenario1}
\end{figure}

One of the modes available to the analysts is to see a historic view of the data as shown in \autoref{fig:Scenario1}. This consists of a visualisation showing four grid views - a Calendar, Daily Top Users, Historic Top Users and Single User Calendar (see \autoref{tab:grids}). The Calendar and Historic Top Users are immediately populated - an analyst can instantly see who caused the most alerts in any 24-hour period over the selected date range (T5), which turns out to be a user with 265 alerts on the 15th March. Selecting this user populates the Single User Calendar - here we can see that this day is the central peak of around a week's (Wed - Wed) worth of alert triggering (T8). Outside this time they never register more than 2 alerts per day.

Selecting the grid item also sends that item's alerts to the facetplot where organising the plot can reveal if these alerts share a common trigger (T4). The 'Show Alert constants' tab reveals these are all related to a cloud-based policy, and selecting attributes 'Resource Type' and 'Resource' reveals all but one of these 265 alerts to be backing up the same powerpoint file to the cloud. In effect, this appears to be someone working on an important file which they have set up to autosave as they work on it. Setting the third facetplot axis, colour, to alert time confirms this - it has been backed up repeatedly between 8:30 am and 3pm (T2). For completeness, we select Friday 12th March and Tuesday 16th March from the Single User Calendar and the same pattern and file is shown in the facetplot. In summary, these alerts were all triggered by one policy on the same file, perhaps a case of over-sensitivity on the policy's part.

\section{Conclusion}

The work here acts as a case study in introducing visualisation capabilities into a commercial product, with a real world dataset providing a realistic foundation for data typing, scale and qualities and with practising security analysts providing feedback.

The final interface provides a multiple view visualisation that can support support the identified tasks T1-8, and also demonstrates how the current product will improve from incorporating more well-designed and appropriate visualisations features into the UI. It also drills deeper into a typical insider threat workflow in that rather than just present filtered sets or instances of events, it allows analysts to find items of interest, then find related users or resources and in turn discover what they are related to.

There are limitations in that this application has been used on a single dataset which leads to the fundamental question 'are the patterns within it typical' e.g. is this a typical number of alerts per user, a typical number of rules established by the analysts, a typical distribution of alerts per day and per rule? However, this is also a real world dataset that has been gathered over 2 years from a large organisation whose userbase and needs are likely to be a superset of a smaller customer. It also focuses on signature-based alerts rather than more contemporary AI generated alerts, as the analysts we interviewed said they trusted their own judgement over a black-boxed algorithm as to what insider actions were potentially damaging to their organisation.

It was interesting that while the tasks we gathered from the requirements proved useful, talking to the analysts provided insight over and above these - most importantly how these tasks were related and connected - for example finding relationships between users or resources always followed on from finding a suspicious user or resource. In effect the tasks we had gathered were pieces of a jigsaw and the analysts feedback helped us recognise how to put them together, further evidence of how involving real users is the most crucial aspect in building an interactive system.

\section {Future Work}

It is clear that the analysts appear wary of AI generated alerts, and some method of overcoming this is necessary. Explainable AI (XAI) \cite{Sec178}, is the practice of increasing trust in and transparency of AI systems by demystifying how decisions are arrived at, and often specifically uses visualisation to do so as seen in the emerging topic of machine learning visualisation \cite{Sec13}. Our premise is that if the AI alerts can be displayed alongside their nearest rule-based alerts - rules which are configured by the analysts - and show meaningful correlations, it could aid analysts in understanding why AI alerts are fired when they are, and thus increasing their confidence in them. The successful fusing of both signature-based and understandable AI-based detection would be a marked step forward.

\acknowledgments{
The authors wish to thank the customer who allowed their data to be used anonymously for the purpose of annotating the screenshots in this paper and for their analysts' co-operation in supplying feedback to visualisation prototypes. Also, we'd like to thank colleagues at FortiNet who helped with setting up back end infrastructure for the visualisation and aided in the understanding of the source data and the FortiInsight product. This work was supported and funded by UK Knowledge Transfer Partnership (KTP) Grant 1025012 along with Edinburgh Napier University and Fortinet Ltd.}

\bibliographystyle{abbrv-doi}

\bibliography{SecurityOnly,Martin9-Saved-Converted}

\end{document}